# Model of acousto-optic diffraction of light in 2-D photonic crystals

Z.A. Pyatakova

119991, Moscow, Leninskie gory, 1, MSU, Faculty of Physics
zoya.pyatakova@physics.msu.ru

The model of nonlinear interaction of proper waves of photonic crystal with plane acoustic wave was developed. The formulation of the model is reduced to the eigenvalue problem, which can be solved by computer simulations. By means of the formulae given in present paper one can predict which polarizations of acoustic wave can result in Bragg diffraction of optical waves of TE or TM polarizations. Computer simulation allows obtaining amplitudes of interaction waves in the case of Bragg diffraction when phase-matching conditions are fulfilled.

## Introduction

Acousto-optic interaction in photonic crystals is now an object of increasing interest [1-3]. Photonic crystals can become an efficient material for acousto-optic devices. There are several mechanisms that allow expecting new effects. First, it is the effect of “slow” light and sound. The figure of merit in acousto-optic diffraction is inverse proportional to sound velocity [4]. That is why small group velocity near the band edge of photonic crystal is of special interest [5]. Another mechanism is complex dispersion law and artificial anisotropy that can lead to some peculiarities in frequency dependences of Bragg angle and can give new geometries of acousto-optic diffraction. All these possible effects now require computation and analysis.

First step of the analysis is to calculate the frequency dependences of Bragg angle, i.e. to analyze the condition of phase matching for photons and phonons in photonic crystal lattice. Such analysis was carried out, for example, in [6], where the method of computation of these dependences was developed.

Second step has to analyze the diffraction efficiency of light. There can be several approaches. First is full-vectorial modeling of propagation of acoustic and optical waves based on solving of wave equations for example by means of FDTD. Second is using the approach of effective media and calculation of effective refractive index, coefficients of stiffness and so on. First approach is very complex and requires much computation time, also it doesn't allow fast variation of parameters and the analysis of results is very complex. Second approach works only then the wavelength of light is much larger than period of photonic crystal lattice and this approximation is out of many interesting cases.

So, the aim of current work is to develop convenient method for modeling of interaction of acoustical and optical waves. We consider the case of Bragg diffraction because it is simpler and more prospective for applications.

## Acousto-optic interaction of proper waves in photonic crystal

Consider the wave equation for vector of magnetic field H in inhomogeneous media:

$$\mathrm{rot}\frac{1}{\varepsilon}\mathrm{rot}\,\mathbf{H} = \frac{1}{c^2}\frac{\partial^2 \mathbf{H}}{\partial t^2} \qquad (1)$$

$$\varepsilon(\mathbf{r},t) = \varepsilon(\mathbf{r}) + \Delta\varepsilon(\mathbf{r},t), \quad \Delta\varepsilon(\mathbf{r},t) = \Delta\varepsilon_0(\mathbf{r})e^{i(\Omega t - \mathbf{Kr})}, \ \Omega \ll \omega, \quad \Delta\varepsilon \ll \varepsilon;$$

Here $\Delta\varepsilon$ is an addition by means of photoelastic effect, $\Omega$ and $\omega$ are sound and light frequencies, respectively, K is the wavevector of sound

As it is done for homogeneous medium, we derive the solution as a sum of harmonic waves with frequencies $\omega + n\Omega$. In the case of Bragg diffraction only two waves are left – incident $H_0$ and diffracted $H_1$. The terms "incident" and "diffracted" are taken for considerations of uniform media, for photonic crystal they are not consistent because the wave which enters the photonic crystal from air experiences the diffraction on photonic crystal lattice. So, in the current paper the term "incident wave" will mean the wave, propagating in photonic crystal before interaction with sound, and "diffracted wave" will mean the wave which appeared from incident wave by means of nonlinear interaction with sound. For these waves we obtain the following system of equations:

$$
\begin{aligned}
&\operatorname{rot}\varepsilon^{-1}\operatorname{rot}\mathbf{H}_0-\operatorname{rot}\Delta\varepsilon_o^{-1}\operatorname{rot}\mathbf{H}_1=\frac{\omega^2}{c^2}\mathbf{H}_0\\
&\operatorname{rot}\varepsilon^{-1}\operatorname{rot}\mathbf{H}_1-\operatorname{rot}\Delta\varepsilon_o^{-1}\operatorname{rot}\mathbf{H}_0=\frac{(\omega-\Omega)^2}{c^2}\mathbf{H}_1
\end{aligned}
\qquad (3)
$$

We know that photonic crystal, not perturbed by acoustic wave, is the periodic medium, and its eigenwaves obey the Bloch theorem [7]. So we try for the solution as a sum of Bloch waves, with the amplitude slowly varies along the direction of interaction as a function *C(x)*

$$
\begin{aligned}
&\mathbf{H_0}(\mathbf{r})=\mathbf{C_0}(x)\sum_{\mathbf{G}}\mathbf{M_0}e^{i(\mathbf{k_0}+\mathbf{G})\mathbf{r}}\\
&\mathbf{H}_1(\mathbf{r})=\mathbf{C}_1(x)\sum_{\mathbf{G}}\mathbf{M}_1e^{i(\mathbf{k}_1+\mathbf{G})\mathbf{r}}
\end{aligned}
\qquad (4)
$$

where $\mathbf{M}_{0,1}$ – amplitudes of Bloch waves, corresponding to incident and diffracted waves, $\mathbf{C}_{0,1}$ – envelopes slowly varying along the spatial coordinate x and independent from other coordinates, $\mathbf{G}$ – reciprocal lattice vector. The suggestion of slow variation allows us to eliminate second derivatives $d^2\mathbf{C}_{0,-1}/dx^2$. Also wee use the phase-matching condition that allows to get rid of spatial variations of Fourier components of inverse dielectric tensor.

Inverse dielectric tensor and its perturbation can also be presented as a Fourier series:

$$
\begin{aligned}
&\frac{1}{\varepsilon(\mathbf{r})}=\sum_{\mathbf{G}}\varsigma(\mathbf{G})e^{i\mathbf{G}\mathbf{r}} && \Delta\varepsilon_o^{-1}(\mathbf{r})=\sum_{\mathbf{G}}\Delta\varsigma(\mathbf{G})e^{i\mathbf{G}\mathbf{r}}\\
&\varsigma(\mathbf{G})=\frac{1}{S}\int_S\frac{1}{\varepsilon(\mathbf{r})}e^{-i\mathbf{G}\mathbf{r}}d^2r, \text{ and} && \Delta\varsigma(\mathbf{G})=\frac{1}{S}\int_S\Delta\varepsilon_o^{-1}(\mathbf{r})e^{-i\mathbf{G}\mathbf{r}}d^2r
\end{aligned}
\qquad (5)
$$

Substituting (4) and (5) into (3) and accounting that $\mathbf{M}_{0,1}$ obey the wave equation with zero perturbation of dielectric constant, we obtain the following system of first order differential equations.

$$
\begin{cases}
\tilde{A}_{10}\dfrac{d\tilde{C}_0}{dx}=\tilde{B}_{20}\tilde{C}_1+\tilde{B}_{10}\dfrac{d\tilde{C}_1}{dx};\\
\tilde{A}_{11}\dfrac{d\tilde{C}_1}{dx}=\tilde{B}_{21}\tilde{C}_0+\tilde{B}_{11}\dfrac{d\tilde{C}_0}{dx}
\end{cases}
\text{где } \tilde{C}_{0,1}=C_{0,1}M_{0,1}
\qquad (6)
$$

The system (6) can be reduced to the generalized eigenvalue problem by means of standard substitution $\begin{bmatrix}\tilde{C}_0\\ \tilde{C}_{-1}\end{bmatrix}=\mathbf{C}_\lambda e^{\lambda x}$:

$$
\begin{pmatrix}\tilde{A}_{10} & -\tilde{B}_{10}\\ -\tilde{B}_{11} & \tilde{A}_{11}\end{pmatrix}\begin{pmatrix}\lambda_1 & 0\\ 0 & \lambda_2\end{pmatrix}\mathbf{C}_\lambda=\begin{pmatrix}0 & \tilde{B}_{20}\\ \tilde{B}_{21} & 0\end{pmatrix}\mathbf{C}_\lambda
\qquad (7)
$$

Matrices included in equations (7) contain all components of dielectric tensor, wavevectors of incident and diffracted waves. In the common case of arbitrary anisotropy every Fourier component of $\mathbf{C}_\lambda$ contains 6 coordinates (3 for incident wave and 3 for diffracted wave). In the case when TE and TM waves are interacting, the matrices simplify, the formulations of eigenvalue problems for these cases are given in Appendix.

**Peculiarities of model and computer simulations**

The view of matrices allows us to predict the existence of acoustooptical diffraction with different crystal symmetries and polarizations of acoustic waves. Acoustic wave can change different components of dielectric tensor, but not all changes lead to diffraction. As we see, for isotropic diffraction of TE and TM waves acoustic wave has to change diagonal components of dielectric tensor. For anisotropic diffraction it is necessary for acoustic wave to change both diagonal and non-diagonal components of dielectric tensor, so the polarization of acoustic wave should have both longitudinal and shear components.

The eigenvalue problem (7) can be solved by means of computer simulations. The size of matrices is determined by the number of plane waves in equation (4) which are taken for computations. The initial conditions in x=0 are unity for the component of incident wave, which corresponds to G=0 and zero for components of diffracted waves and other components of incident wave. The solving of equation (7) has much in common with the calculation of band structures for photonic and phononic crystal by means of plane-wave expansion method (PWE, see, for example, in [8]). The number of plane waves can be chosen the same as most common cases of band structures. The computational problem that can be faced is violation of energy conservation law, because eigenvalues may have nonzero real part. In the case of uniform media all eigenvalues are pure imagine, so it gives oscillating solution along x axis, which is fully described by existing theory [4]. Positive real part in eigenvalue of diffracted wave without corresponding negative part for incident wave can give exponentially growing solutions, so we have to pay much attention to them. The restriction to number of interacting harmonics or length of interaction allows minimize the instability of numerical scheme.

**Simulation results and discussion**

The example of numerical modeling is shown at figure 1. Here is the simultaneous calculation of Bragg angle and diffraction efficiency (amplitude of diffracted wave) for the given normalized frequency of light and sound (normalized frequency is ratio of lattice period to wavelength). For calculation we considered the model system: silicon matrix with periodically implemented cylindrical silica rods, forming the square lattice. We considered the isotropic diffraction of TM polarizations of light, normalized frequency of light is 0.2 (i.e. for lattice period of 1 μm wavelength is 5 μm). Sound has longitudinal (L) polarization; amplitude displacement in acoustic wave is 1 nm.

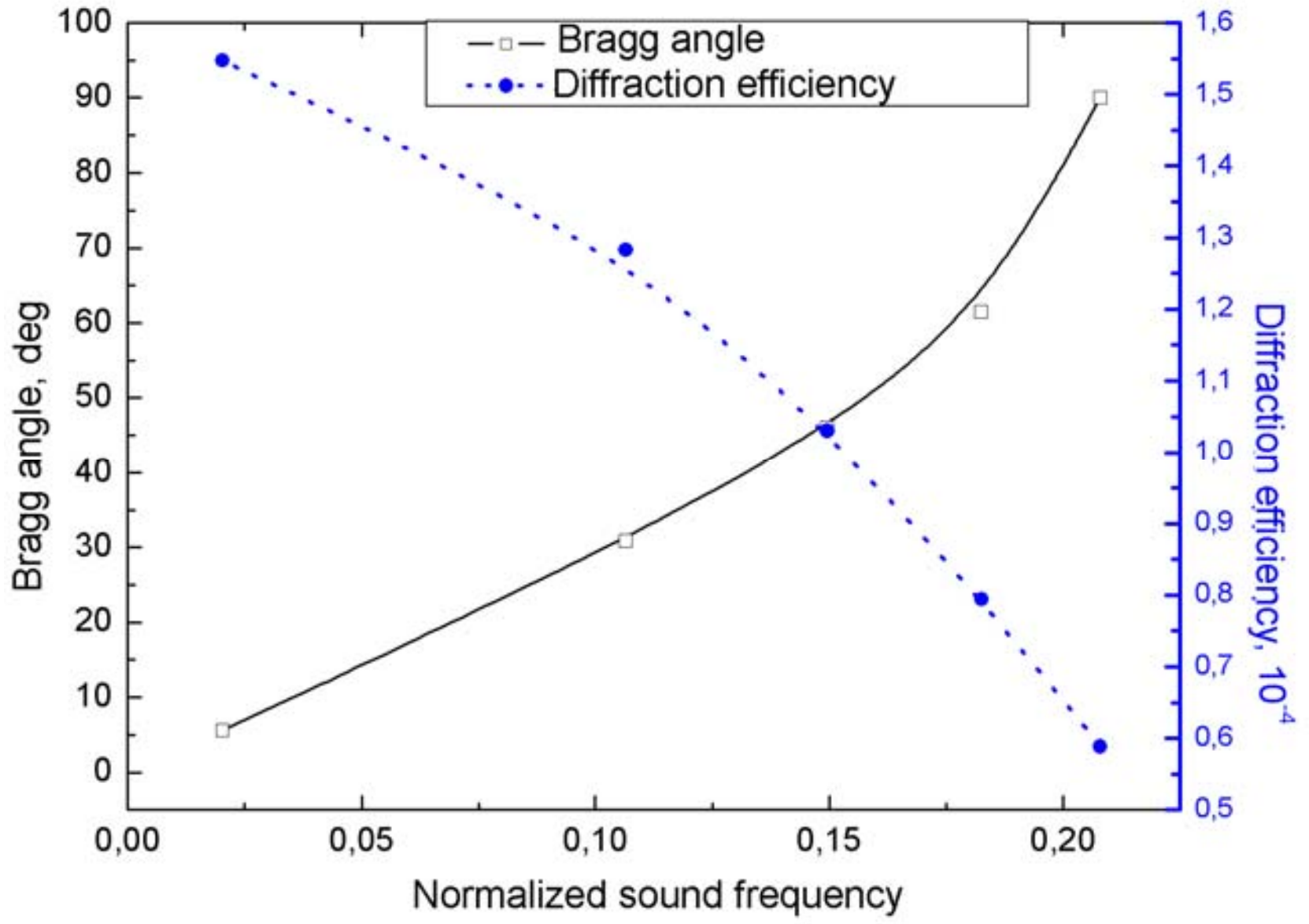

Fig 1. Example of numerical simulation of acoustooptic interaction in photonic crystal: black curve is Bragg angle, defined by phase-matching condition add blue curve is diffraction efficiency (amplitude of diffracted wave), diffraction length is 500 μm, incident ad diffracted light polarization is TM, sound polarization is L, normalized frequency of light is 0.2. Photonic crystal is silica rods in silicon matrix, square lattice. Amplitude of acoustic wave is 1 nm.

The test of proposed model was conducted on uniform media when dielectric permittivity and elastic constants of both components of photonic crystal are the same. The calculations with the material parameters of $\alpha$-$SiO_2$ demonstrate the correspondence with existing theoretical and experimental results.

Note that diffraction efficiency calculated for considered system is not higher than in homogeneous silicon and silica. Figure 2 shows the dependence of amplitudes on diffraction length for different filling factors, other parameters are the same as in fig. 1, normalized frequency of sound is 0.20.

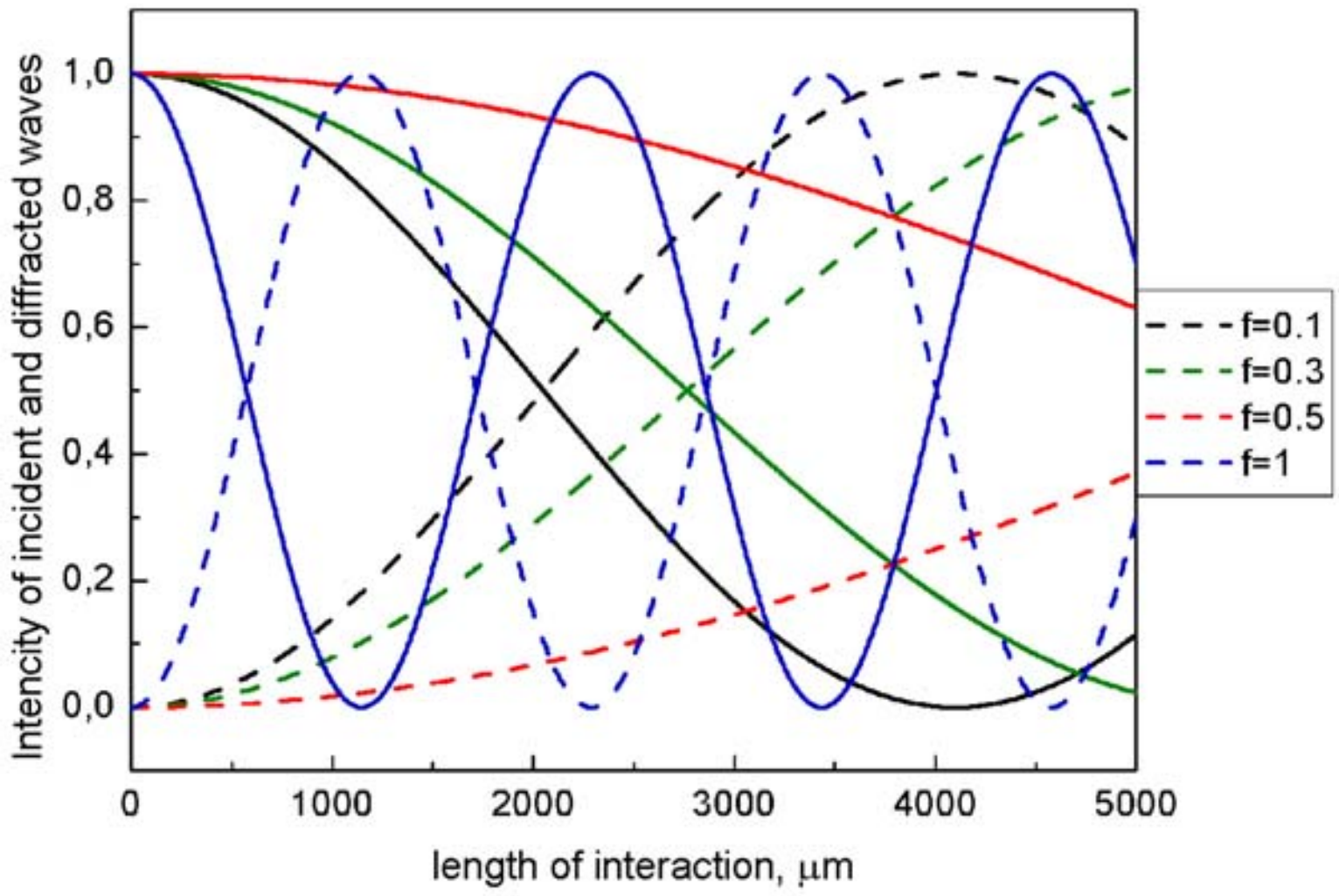

Fig.2. Spatial dependence of intensity of incident (solid lines) and diffracted (dashed lines) optical waves in photonic crystals with different filling factors. Other calculation parameters are the same as in fig. 1.

So, it is necessary to optimize the geometry of interaction for obtaining better results. As we can see, photonic crystals have a lot of parameters which can be varied.

**Conclusion**

The method developed in the present paper can be useful for understanding the problem of acoustooptic interaction in photonic crystals and designing new photonic crystal devices of acoustooptics. I would like to thank G.V. Belokopytov and A.P. Pyatakov for the discussions of proposed model.

**Appendix.** Matrices for calculation of diffraction efficiency in the case of TE and TM waves.

$\mathbf{A}\frac{d\tilde{\mathbf{C}}}{dx}=\mathbf{B}\tilde{\mathbf{C}}$ – differential equation. $\tilde{C}_0$ и $\tilde{C}_1$ – amplitudes of incident and diffracted waves $\varsigma_{ij},\Delta\varsigma_{ij}$ – Fourier components of inverse dielectric tensor and their perturbed part, $G,G'$ are independent sets of reciprocal lattice vectors. Matrices **A** and **B** are presented in the table.

| **A** | **B** |
|---|---|
| Isotropic, TE-TE, $\tilde{C}_0^{\,x},\tilde{C}_0^{\,y},\tilde{C}_1^{\,x}\tilde{C}_1^{\,y}$ | |
| $\begin{pmatrix} 0 & \varsigma_{33}(k_{0y}+G_y') & 0 & -\Delta\varsigma_{33}(k_{1y}+G_y') \\ \varsigma_{33}(k_{0x}+G_x') & 0 & -\Delta\varsigma_{33}(k_{1x}+G_x') & 0 \\ 0 & \Delta\varsigma_{33}(k_{0y}+G_y') & 0 & \varsigma_{33}(k_{1y}+G_y') \\ -\Delta\varsigma_{33}(k_{0x}+G_x') & 0 & \varsigma_{33}(k_{1y}+G_y') & 0 \end{pmatrix}$ | $\begin{pmatrix} 0 & 0 & \Delta\varsigma_{33}(k_{1y}+G_y)\cdot(k_{-1y}+G_y') & -\Delta\varsigma_{33}(k_{1x}+G_x)\cdot(k_{-1y}+G_y') \\ 0 & 0 & -\Delta\varsigma_{33}(k_{1x}+G_x')\cdot(k_{1y}+G_y) & \Delta\varsigma_{33}(k_{1x}+G_x)\cdot(k_{1x}+G_x') \\ -\Delta\varsigma_{33}(k_{0y}+G_y)\cdot(k_{0y}+G_y') & \Delta\varsigma_{33}(k_{0x}+G_x)\cdot(k_{0y}+G_y') & 0 & 0 \\ \Delta\varsigma_{33}(k_{0y}+G_y)\cdot(k_{0x}+G_x') & -\Delta\varsigma_{33}(k_{0x}+G_x)\cdot(k_{0x}+G_x') & 0 & 0 \end{pmatrix}$ |
| Isotropic, TM-TM, $\tilde{C}_0^{\,z},\tilde{C}_1^{\,z}$ | |
| $\begin{pmatrix} -\varsigma_{12}((k_{0y}+G_y)+(k_{0y}+G_y'))+\varsigma_{22}((k_{0x}+G_x)+(k_{0x}+G_x')) & -\Delta\varsigma_{12}((k_{-1y}+G_y)+(k_{-1y}+G_y'))+\Delta\varsigma_{22}((k_{-1x}+G_x)+(k_{-1x}+G_x')) \\ \Delta\varsigma_{12}((k_{0y}+G_y)+(k_{0y}+G_y'))-\Delta\varsigma_{22}((k_{0x}+G_x)+(k_{0x}+G_x')) & \varsigma_{12}((k_{-1y}+G_y)+(k_{-1y}+G_y'))-\varsigma_{22}((k_{-1x}+G_x)+(k_{-1x}+G_x')) \end{pmatrix}$ | $\begin{pmatrix} 0 & -\Delta\varsigma_{11}(k_{-1y}+G_y)(k_{-1y}+G_y')-\Delta\varsigma_{22}(k_{-1x}+G_x)(k_{-1x}+G_x')+\Delta\varsigma_{12}(k_{-1x}+G_x)(k_{-1y}+G_y')+\Delta\varsigma_{12}(k_{-1x}+G_x')(k_{-1y}+G_y) \\ -\Delta\varsigma_{11}(k_{0y}+G_y)(k_{0y}+G_y')-\Delta\varsigma_{22}(k_{0x}+G_x)(k_{0x}+G_x')+\Delta\varsigma_{12}(k_{0x}+G_x)(k_{0y}+G_y')+\Delta\varsigma_{12}(k_{0x}+G_x')(k_{0y}+G_y) & 0 \end{pmatrix}$ |
| Anisotropic, TE-TM, $\tilde{C}_0^{\,x},\tilde{C}_0^{\,y},\tilde{C}_1^{\,z}$ | |
| $\begin{pmatrix} 0 & \varsigma_{33}(k_{0y}+G_y') & -\Delta\varsigma_{23}(k_{1y}+G_y') \\ \varsigma_{33}(k_{0x}+G_x') & 0 & \Delta\varsigma_{23}((k_{1x}+G_x)+(k_{0x}+G_x')) \\ 0 & 0 & \Delta\varsigma_{12}((k_{1y}+G_y)+(k_{1y}+G_y'))-\Delta\varsigma_{22}((k_{1x}+G_x)+(k_{1x}+G_x')) \end{pmatrix}$ | $\begin{pmatrix} 0 & 0 & \Delta\varsigma_{13}(k_{1y}+G_y)\cdot(k_{1y}+G_y')-\Delta\varsigma_{23}(k_{1x}+G_x)\cdot(k_{1x}+G_y') \\ 0 & 0 & -\Delta\varsigma_{13}(k_{1y}+G_y)\cdot(k_{1y}+G_x')+\Delta\varsigma_{23}(k_{1x}+G_x)\cdot(k_{1x}+G_x') \\ \Delta\varsigma_{33}(k_{0y}+G_y)\cdot(k_{1y}+G_y') & \Delta\varsigma_{33}(k_{0y}+G_y)\cdot(k_{1x}+G_x') & -\Delta\varsigma_{11}(k_{1y}+G_y)(k_{1y}+G_y')-\Delta\varsigma_{22}(k_{1x}+G_x)(k_{1x}+G_x')+\Delta\varsigma_{12}(k_{1x}+G_x)(k_{1y}+G_y')+\Delta\varsigma_{12}(k_{1x}+G_x')(k_{1y}+G_y) \end{pmatrix}$ |
| Anisotropic, TM-TE, $\tilde{C}_0^{\,z},\tilde{C}_1^{\,x},\tilde{C}_1^{\,y}$ | |
| $\begin{pmatrix} \varsigma_{12}((k_{0y}+G_y)+(k_{0y}+G_y'))-\varsigma_{22}((k_{0x}+G_x)+(k_{0x}+G_x')) & 0 & 0 \\ 0 & 0 & \varsigma_{33}(k_{-1y}+G_y') \\ 0 & \varsigma_{33}(k_{-1x}+G_x') & 0 \end{pmatrix}$ | $\begin{pmatrix} 0 & \Delta\varsigma_{33}(k_{-1y}+G_y)\cdot(k_{1y}+G_y') & -\Delta\varsigma_{33}(k_{-1x}+G_x)\cdot(k_{1y}+G_y') \\ 0 & -\Delta\varsigma_{33}(k_{1x}+G_x')\cdot(k_{1y}+G_y) & \Delta\varsigma_{33}(k_{1x}+G_x)\cdot(k_{1x}+G_x') \\ -\Delta\varsigma_{11}(k_{0y}+G_y)(k_{1y}+G_y')-\Delta\varsigma_{22}(k_{0x}+G_x)(k_{1x}+G_x')+\Delta\varsigma_{12}(k_{0x}+G_x)(k_{1y}+G_y')+\Delta\varsigma_{12}(k_{0y}+G_y)(k_{1x}+G_x') & 0 & 0 \end{pmatrix}$ |